\documentstyle[prl,aps,twocolumn]{revtex}

\begin{document}

\title{
{\normalsize\begin{flushright}
UB--ECM--PF--97/12\\
hep-th/9707063\\[1.5ex]
\end{flushright}}
The rigid symmetries of bosonic D-strings
}
\author{
Friedemann Brandt, Joaquim Gomis, Joan Sim\'on}

\address{
Departament ECM,
Facultat de F\'{\i}sica,
Universitat de Barcelona and 
Institut de F\'{\i}sica d'Altes Energies,
Diagonal 647,
E-08028 Barcelona, Spain. E-mail addresses:
brandt@ecm.ub.es, gomis@ecm.ub.es, jsimon@ecm.ub.es.
\\[1.5ex]
\begin{minipage}{14cm}\rm\quad
We analyse the classical symmetries of bosonic D-string actions
and generalizations thereof. Among others, we show
that the simplest actions of this type have infinitely many
nontrivial rigid symmetries which act nontrivially and
nonlinearly both on the target space coordinates 
and on the $U(1)$ gauge field, and form a Ka\v{c}-Moody version
of the Weyl algebra (= Poincar\'e algebra + dilatations).\\[1ex]
\end{minipage}
}

\maketitle

\section*{Introduction}

Much progress has been made lately in constructing 
$\kappa$-invariant actions
for $D$-$p$-branes \cite{actions1,actions2,actions3},
 generalizing earlier work \cite{actions}.
Typically, the ``bosonic part'' of these actions is of the
Born--Infeld type,
such as
\begin{eqnarray}
S_p&=&\int d^{p+1}\sigma
\sqrt {|\det({\cal G}_{\mu\nu}+F_{\mu\nu})|}\ ,
\nonumber\\
{\cal G}_{\mu\nu}&=&\eta_{mn}\partial_\mu x^m \cdot \partial_\nu x^n,
\nonumber\\
F_{\mu\nu}&=&\partial_\mu a_\nu-\partial_\nu a_\mu\ .
\label{Sp}
\end{eqnarray}
Here, the $x^m$ are target space coordinates,
$\eta_{mn}$ is a flat target space metric,
and $a_\mu$ is an abelian gauge field living in the
world volume.

Important properties of actions are of course
their symmetries. In particular one may ask:
What are the rigid and gauge symmetries of
(\ref{Sp})? To what
degree is this action determined by symmetries
alone? In this letter we analyse these questions for
the case $p=1$. Our results apply not only to
the action (\ref{Sp}) but also to generalizations thereof which
will be given below. We obtained these results by an
analysis of the BRST cohomology which will be given in \cite{inprep}.

Although the cohomological analysis parallels quite closely
the one for bosonic strings carried out in \cite{paper1}, 
the results are surprisingly
rather different. For instance, while the usual bosonic string
in a flat target space has only finitely many rigid symmetries
before gauge fixing \cite{paper1}, we
will show that, for $p=1$, the action (\ref{Sp}) has
{\em infinitely many} nontrivial rigid symmetries. Among them 
there are of course the obvious Poincar\'e symmetries which
reflect the isometries of the
target space and coincide with the 
rigid symmetries of the bosonic string. 
However we find also previously unnoticed rigid symmetries which
are nonlinearly realized and transform both the $x^m$ and 
the gauge field $a_\mu$. Together with the familiar
Poincar\'e symmetries, the new symmetries form a Ka\v{c}--Moody version
of the Weyl algebra (= Poincar\'e algebra + dilatations).
We stress that these symmetries are present already 
{\em before} fixing a gauge. They should
therefore not be confused with the Ka\v{c}--Moody symmetries 
of sigma models discussed e.g.\ in \cite{hull} as the latter
emerge just as residual symmetries of Weyl and diffeomorphism
invariant actions in suitable gauges.

The fact that the new symmetries transform $a_\mu$ nontrivially
has remarkable consequences. In particular there are 
symmetry transformations which map
solutions of the equations of motion with trivial 
gauge field (zero or pure gauge) to other solutions
with nonvanishing field strength $F_{\mu\nu}$, 
and thus usual bosonic strings to
$D$-strings. As the field strength
contributes to the string tension \cite{schwarz}, the
new symmetries therefore also relate strings with different tension
and might thus be viewed as ``stringy symmetries''.
It is striking that these properties of the new symmetries
are similar to those of dualities \cite{tseytlin,dual}
relating bosonic and $D$-strings. One may speculate
whether the new symmetries reflect part of the
symmetry structure of an underlying (``M'') theory. As they
are nonlinearly realized, one might for instance suspect
that they emerge somehow as broken symmetries of that theory.

\section*{Actions}

To motivate and explain our approach, we note that
(\ref{Sp}) can be cast in a more
convenient form \cite{lind,hull1} with
Lagrangian
\begin{eqnarray}
L_p=\frac 12\, \sqrt{\varrho}\, [ \varrho^{\mu\nu}
({\cal G}_{\mu\nu}+F_{\mu\nu})-(p-1)]
\label{PSp}
\end{eqnarray}
where $\varrho=|\det(\varrho_{\mu\nu})|$.
In this formulation, the $\varrho_{\mu\nu}$ are
auxiliary fields ($\varrho^{\mu\nu}$
denotes the inverse of $\varrho_{\mu\nu}$). Eliminating 
them, one recovers (\ref{Sp}). 

The action with Lagrangian
(\ref{PSp}) is evidently invariant under world-volume
diffeomorphisms and abelian gauge transformations
of $a_\mu$. For $p=1$ it is in addition gauge invariant
under Weyl transformations of $\varrho_{\mu\nu}$
and we can decompose the latter according to
\begin{equation}
p=1:\ \varrho_{\mu\nu}=
\gamma_{\mu\nu}+\sqrt {\gamma}\, \epsilon_{\mu\nu}\varphi
\label{f}
\end{equation}
with 
\[\gamma_{\mu\nu}=\gamma_{\nu\mu},\
\gamma=-\det(\gamma_{\mu\nu}),\ 
\epsilon_{21}=\epsilon^{12}=1\] 
where we assumed for definiteness that $\gamma_{\mu\nu}$ has Lorentzian
signature.
Since $\sqrt {\gamma}\,\epsilon_{\mu\nu}$ behaves as a covariant
2-tensor field under
world-sheet diffeomorphisms and has the same Weyl weight
as $\gamma_{\mu\nu}$,
$\varphi$ transforms as a scalar
field under diffeomorphisms and is Weyl invariant.
Using (\ref{f}), the Lagrangian (\ref{PSp}) for $p=1$ reads
\begin{equation}
L_1=\frac 12\, (1-\varphi^2)^{-1/2}
(\sqrt{\gamma}\, \gamma^{\mu\nu}{\cal G}_{\mu\nu}
-\varphi\,\epsilon^{\mu\nu}  F_{\mu\nu}).
\label{PS1}
\end{equation}

We are now looking for generalizations of (\ref{PS1}), guided
by its field content and gauge symmetries.
Since the gauge symmetries of (\ref{PS1}) treat $\varphi$ on
an equal footing with the $x^m$, we can treat $\varphi$ has zeroth
coordinate of an extended target space with coordinates $X^M$,
\[ \{X^M\}=\{\varphi\, ,\ x^m\},\quad \varphi\equiv X^0.\]
Furthermore we allow for a set of abelian gauge fields $a^I_\mu$
rather than only one such gauge field, and
fix the field content of the models to be studied to
\begin{equation}
\{\phi^i\}=\{\gamma_{\mu\nu}\, ,\ a^I_\mu\, ,\ X^M\}.
\label{fields}
\end{equation}
In addition to this field content, we impose
gauge invariance under world-sheet diffeomorphisms,
Weyl transformations of the $\gamma_{\mu\nu}$,
and abelian gauge transformations of the $a^I_\mu$.
Infinitesimally these gauge transformations read
\begin{eqnarray}
\delta \gamma_{\mu\nu}&=&
\varepsilon^\rho\partial_\rho \gamma_{\mu\nu}
+2 \gamma_{\rho(\nu}\partial_{\mu)}\varepsilon^\rho
+\lambda \gamma_{\mu\nu}\ ,
\nonumber\\
\delta a^I_\mu&=&
\varepsilon^\nu\partial_\nu a^I_\mu 
+a^I_\nu\partial_\mu\varepsilon^\nu+\partial_\mu \Lambda^I\ ,
\nonumber\\
\delta X^M &=&\varepsilon^\mu\partial_\mu X^M
\label{brs}
\end{eqnarray}
where $\varepsilon^\mu$, $\lambda$ and $\Lambda^I$ parametrize
world-sheet diffeomorphisms, Weyl transformations and 
gauge transformations of the $a_\mu^I$ respectively. 
Our first result is that, up to a total derivative,
the most general Lagrangian which is
(a) constructible solely of the fields (\ref{fields}),
(b) local (= polynomial in derivatives of any order),
and (c) up to a total derivative invariant
under the gauge transformations (\ref{brs}), is
\begin{eqnarray}
L &=&
\frac 12\, \sqrt \gamma\, \gamma^{\mu\nu}G_{MN}(X)\,
\partial_\mu X^M\cdot\partial_\nu X^N
\nonumber\\
& &{}
+\epsilon^{\mu\nu}[\, \frac 12\, B_{MN}(X)\,
\partial_\mu X^M\cdot\partial_\nu X^N
\nonumber\\
& &{}
+D_I(X)\, \partial_\mu a^I_\nu\, ]
\label{L}
\end{eqnarray}
where the $G_{MN}$, $B_{MN}$ and $D_I$ are arbitrary functions
of the $X^M$. This result 
is the analogue of a similar one holding for
bosonic strings \cite{paper1} and will be proved
in \cite{inprep}.
Note that (\ref{L}) covers in particular
$D$-string actions of a general form, if we choose
\begin{eqnarray}
& & G_{M0}=0,\ G_{mn}=g_{mn}(x)\, f(\varphi),\
\nonumber\\
& & B_{m0}=0,\ B_{mn}=b_{mn}(x)\sqrt{f^2(\varphi)-1} ,
\nonumber\\
& & D_I=d_I(x)\sqrt{f^2(\varphi)-1}
\label{BI1}\end{eqnarray}
where $f(\varphi)$ is (almost) arbitrary (this 
arbitrariness reflects the freedom of field redefinitions
$\varphi\rightarrow\tilde\varphi(\varphi)$).
Indeed, upon elimination of
$\gamma_{\mu\nu}$ and $\varphi$,
(\ref{BI1}) yields Born--Infeld actions
generalizing (\ref{Sp}) among others to curved target spaces:
\begin{eqnarray}
S_1&=&\int d^2\sigma
\sqrt {-\det(\hat {\cal G}_{\mu\nu}+{\cal F}_{\mu\nu})},
\nonumber\\
\hat {\cal G}_{\mu\nu}&=&
g_{mn}(x)\,\partial_\mu x^m \cdot \partial_\nu x^n,
\nonumber\\
{\cal F}_{\mu\nu}&=&
d_I(x)\,(\partial_\mu a^I_\nu-\partial_\nu a^I_\mu)
\nonumber\\
& &+b_{mn}(x)\, \partial_\mu x^m \cdot \partial_\nu x^n.
\label{BI2}
\end{eqnarray}

We note that (\ref{L}) covers for instance also actions with
Lagrangian
\begin{equation}
L=\sqrt {-\det(\hat {\cal G}_{\mu\nu}+{\cal F}_{\mu\nu})}
-\sqrt{-\det(\hat {\cal G}_{\mu\nu})}
\label{BI3}\end{equation}
which were considered already by Born and Infeld 
(for $p=3$) \cite{BI}
and are obtained analogously by choosing
$G_{mn}=g_{mn}(x)\, [f(\varphi)-1]$ and the 
remaining functions as in (\ref{BI1}).

\section*{Rigid symmetries}

Our second result is that the nontrivial rigid symmetries of an action
with Lagrangian (\ref{L}) are generated by transformations
\begin{eqnarray}
\Delta X^M&=&{\cal X}^M(X),\quad
\Delta \gamma_{\mu\nu}=0,
\nonumber\\
\Delta a^I_\mu &=&
-\sqrt \gamma\, \epsilon_{\mu\nu}
{\cal A}^I_M(X)\partial^\nu X^M
\nonumber\\
& &
+{\cal B}^I_M(X)\partial_\mu X^M+a^J_\mu {{\cal C}_J}^I(X)
\label{trafos}
\end{eqnarray}
with functions
${\cal X}^M(X)$, ${\cal A}^I_M(X)$, 
${\cal B}^I_M(X)$, ${{\cal C}_J}^I(X)$ solving
\begin{eqnarray}
{\cal L}_{\cal X} G_{MN}&=& -2{\cal A}^I_{(M}\partial_{N)}D_I\ ,
\label{kill1}\\
{\cal L}_{\cal X} B_{MN}&=& 
2\partial_{[N}{\cal Y}_{M]}+2{\cal B}^I_{[N}\partial_{M]}D_I\ ,
\label{kill2}\\
{\cal L}_{\cal X} \partial_M D_I&=&-{{\cal C}_I}^J\partial_M D_J
\label{kill3}\end{eqnarray}
for some functions ${\cal Y}_M(X)$.
Here ${\cal L}_{\cal X}$ denotes the standard Lie derivative along
${\cal X}$, and we used 
\[ 
\partial^\mu=\gamma^{\mu\nu}\partial_\nu\, ,\quad
\partial_M=\partial/\partial X^M.
\]

Note that equations (\ref{kill1}--\ref{kill3}) generalize the familiar 
Killing vector equations for the target space. The general
form of the latter (with nonvanishing $B_{MN}$) was discussed
in \cite{hull,paper1} and arises from (\ref{kill1}--\ref{kill3})
for $D_I=0$ 
(as $D_I=0$ reproduces the usual bosonic string, we thus
recover for this case the result of \cite{paper1} for the rigid
symmetries of the bosonic string).
We will solve these equations explicitly
for specific models in the next section.
In \cite{inprep} we will prove that the
above symmetries exhaust the nontrivial rigid symmetries
of an action with Lagrangian (\ref{L})\footnote{The 
rigid symmetries are obtained from
the BRST cohomology at ghost number $-1$
\cite{bbh1}.}.
Here we only note that, under transformations
(\ref{trafos}) satisfying (\ref{kill1}--\ref{kill3}), the Lagrangian
(\ref{L}) indeed transforms into a total derivative
as can be easily verified,
\begin{eqnarray}
\Delta L &=& \epsilon^{\mu\nu}\partial_\mu  (-{\cal Y}_M\partial_\nu X^M
\nonumber\\
& &
+a^I_\nu\, {\cal X}^M\partial_M D_I+D_I\, \Delta a^I_\nu).
\label{K}
\end{eqnarray}
The conserved Noether currents  $j^\mu$  corresponding to the
symmetries (\ref{trafos}) are now readily computed,
\begin{eqnarray}
j^\mu
&=& \sqrt\gamma\, \gamma^{\mu\nu}G_{MN}{\cal X}^M\partial_\nu X^N
\nonumber\\
& &+\epsilon^{\mu\nu}(\tilde{\cal Y}_M\partial_\nu X^M
-a^I_\nu {\cal X}^M\partial_M D_I)
\label{j}
\end{eqnarray}
where
\[ \tilde{\cal Y}_M={\cal Y}_M-B_{MN}{\cal X}^N.\]

In order to complete the above statements about the rigid symmetries, we
note that the solutions to the generalized Killing vector equations
(\ref{kill1}--\ref{kill3}) are determined only up to the redefinitions
\begin{eqnarray}
{\cal A}^I_M & \rightarrow & {\cal A}^I_M+{\cal E}^{[IJ]}\partial_M D_J\ ,
\nonumber\\
{\cal B}^I_M & \rightarrow & {\cal B}^I_M
+\partial_M {\cal B}^I+{\cal E}^{(IJ)}\partial_M D_J\ ,
\nonumber\\
{\cal Y}_M & \rightarrow & {\cal Y}_M+\partial_M {\cal Y}
-{\cal B}^I \partial_M D_I
\label{triv}
\end{eqnarray}
where the ${\cal B}^I(X)$, ${\cal E}^{IJ}(X)$ and 
${\cal Y}(X)$ are arbitrary functions
of the $X^M$. These redefinitions drop 
out of (\ref{kill1}--\ref{kill3})
and affect in (\ref{trafos})
only the transformations of the $a^I_\mu$ according to 
\begin{eqnarray}
\Delta a^I_\mu \rightarrow  \Delta a^I_\mu
+\partial_\mu {\cal B}^I
+{\cal E}^{IJ}_{\mu\nu}\epsilon^{\nu\rho}\partial_\rho D_J\ ,
\nonumber\\
{\cal E}^{IJ}_{\mu\nu}=
-\sqrt{1/\gamma}\, \gamma_{\mu\nu}{\cal E}^{[IJ]}
+\epsilon_{\mu\nu}{\cal E}^{(IJ)}.
\label{triv1}
\end{eqnarray}
These are irrelevant redefinitions of the rigid symmetries, i.e.\ two
rigid symmetries are identified if they coincide up to such
redefinitions. Namely
$\partial_\mu {\cal B}^I$ are just special gauge transformations,
while ${\cal E}^{IJ}_{\mu\nu}\epsilon^{\nu\rho}\partial_\rho D_J$ 
are on-shell trivial symmetries. The latter holds due to
${\cal E}^{IJ}_{\mu\nu}=-{\cal E}^{JI}_{\nu\mu}$ and
$\epsilon^{\nu\rho}\partial_\rho D_J=\delta S/\delta a^J_\nu$
where $S$ denotes the action with Lagrangian (\ref{L}).

It is easy to check that the commutator of two symmetries
(\ref{trafos}) is again a symmetry of this type,
\begin{equation}
[\Delta_1,\Delta_2]=\Delta_3\ .
\label{comm}\end{equation}
Namely, using (\ref{trafos}) and the notation 
$\Delta_i X^M={\cal X}^M_i$ etc. (i=1,2,3), 
a direct computation
of $[\Delta_1,\Delta_2]$ yields
\begin{eqnarray}
{\cal X}^M_3&=&{\cal X}^N_1\partial_N{\cal X}^M_2-(1\leftrightarrow 2),
\nonumber\\
{\cal A}^I_{3M}&=&{\cal L}_{{\cal X}_1}{\cal A}^I_{2M}
-{{\cal C}_{1J}}^I{\cal A}^J_{2M}-(1\leftrightarrow 2),
\nonumber\\
{\cal B}^I_{3M}&=&{\cal L}_{{\cal X}_1}{\cal B}^I_{2M}
-{{\cal C}_{1J}}^I{\cal B}^J_{2M}-(1\leftrightarrow 2),
\nonumber\\
{{\cal C}_{3J}}^I&=&
{\cal L}_{{\cal X}_1}{{\cal C}_{2J}}^I-(1\leftrightarrow 2).
\label{123}
\end{eqnarray}
Using standard properties of Lie derivatives such as
$[{\cal L}_{{\cal X}_1},{\cal L}_{{\cal X}_2}]=
{\cal L}_{{\cal X}_3}$, it is easy to verify that the 
set of functions (\ref{123}) solves (\ref{kill1}--\ref{kill3}) with
${\cal Y}_{3M}={\cal L}_{{\cal X}_1}{\cal Y}_{2M}
-{\cal L}_{{\cal X}_2}{\cal Y}_{1M}$
whenever the sets
$({\cal X}^M_i,{\cal A}^I_{iM},{\cal B}^I_{iM},
{{\cal C}_{iJ}}^I,{\cal Y}_{iM})$,
$i=1,2$ solve (\ref{kill1}--\ref{kill3}) too. In that sense,
the commutators of symmetry transformations (\ref{trafos}) `close'. 
However, this does {\em not} necessarily imply
that the algebra of the rigid symmetries 
closes off-shell in a particular basis of the rigid symmetries. 
Namely suppose that $\Delta_1$ and $\Delta_2$ are two elements
of such a basis. Their commutator (\ref{comm}) will in general
be a linear combination of elements of the basis only up to
redefinitions (\ref{triv1}). Hence, in general the algebra of the
elements of the basis will close only up to gauge transformations
and on-shell trivial transformations of the type occurring in 
(\ref{triv1}).

To summarize, the nontrivial (infinitesimal) 
rigid symmetries of an action with Lagrangian (\ref{L}) are
exhausted by transformations (\ref{trafos}) with
target space functions satisfying (\ref{kill1}--\ref{kill3}), and defined 
modulo the redefinitions (\ref{triv}). 
Furthermore, any solution to (\ref{kill1}--\ref{kill3}) which
does not vanish modulo redefinitions (\ref{triv}) gives
rise to a nontrivial rigid symmetry generated
by (\ref{trafos}). Hence, one has
precisely to solve (\ref{kill1}--\ref{kill3}) in order to find
all rigid symmetries of an action with Lagrangian (\ref{L}).
A basis of the rigid symmetries is obtained from a basis
of solutions to (\ref{kill1}--\ref{kill3}), i.e.\ from a complete set of
solutions which are linearly independent
up to redefinitions (\ref{triv}).
Needless to say that, on general grounds, rigid symmetries
of Born--Infeld actions (\ref{BI2}) arise from those of
the corresponding actions with Lagrangians 
(\ref{L},\ref{BI1}) by
replacing the auxiliary fields $\gamma_{\mu\nu}$ and 
$\varphi$ in $\Delta x^m$ and $\Delta a^I_\mu$ 
with a solution to
their algebraic equations of motion, i.e.\ by substituting
for instance
\begin{eqnarray}
f(\varphi)&\rightarrow& 
\sqrt {\det(\hat {\cal G}_{\mu\nu})/
\det(\hat {\cal G}_{\mu\nu}+{\cal F}_{\mu\nu})}\ ,
\nonumber\\
\gamma_{\mu\nu}&\rightarrow&\hat {\cal G}_{\mu\nu}\ .
\label{eom}\end{eqnarray}
Note that $\gamma_{\mu\nu}$ is actually defined
by the equations of motion only up to a completely arbitrary function
multiplying $\hat {\cal G}_{\mu\nu}$ due to the Weyl invariance
of (\ref{L}). As this general function drops out of
the transformations (\ref{trafos}), one can indeed choose
(\ref{eom}) with no loss of generality.

\section*{Examples and discussion}

To illustrate and interpret the general results presented
above, we will now solve 
(\ref{kill1}--\ref{kill3}) explicitly for a specific class of 
models and discuss the corresponding rigid symmetries. The models are
characterized by Lagrangians (\ref{L}) involving
only one $U(1)$ gauge field $a_\mu$, with
\begin{eqnarray}
G_{0M}=0,\ G_{mn}=f(\varphi)\,\eta_{mn},\quad\quad
\nonumber\\
B_{0m}=0,\ B_{mn}=B_{mn}(\varphi),\ D=D(\varphi).
\label{simple}
\end{eqnarray}
Recall that the special choice $B_{mn}=0$, $D=(f^2-1)^{1/2}$
reproduces the action (\ref{Sp}) for $p=1$.
We will first show that, for any choice of
$f$ and $D\neq constant$, the general solution
of eqs. (\ref{kill1}--\ref{kill3}) is, 
up to redefinitions (\ref{triv}),
\begin{eqnarray}
{\cal X}^0&=&{\cal X}^0(\varphi),
\nonumber\\
{\cal X}^m&=&- (f'/2f) {\cal X}^0 x^m
\nonumber\\
& &+a^m(\varphi)+a^{mn}(\varphi)x_n,\ a^{mn}=-a^{nm},
\nonumber\\
{\cal A}_m&=&-\eta_{mn} (f/D') ({\cal X}^n)',\
{\cal A}_0= 0,
\nonumber\\
{\cal B}_m&=&(1/D')[B'_{mn}{\cal X}^n+\frac 12(B'_{mn}{\cal X}^0)'x^n],\
{\cal B}_0=0,
\nonumber\\
{\cal C}&=&-({\cal X}^0D')'/D',
\nonumber\\
{\cal Y}_m&=&\frac 12B'_{mn}{\cal X}^0x^n+B_{mn}{\cal X}^n,\ {\cal Y}_0 =0,
\label{sol}
\end{eqnarray}
where ${\cal X}^0(\varphi)$, $a^m(\varphi)$ and $a^{mn}(\varphi)$
are {\em arbitrary functions} of $\varphi$ and we used
\[
 '\equiv\partial/\partial\varphi\ ,\quad x_m\equiv \eta_{mn}x^n\ .
\]
Let us now sketch the derivation of (\ref{sol}).
The results for ${\cal X}^0$ and ${\cal C}$ follow immediately from
(\ref{kill3}) as it reads in the cases under study
$D'\partial_m {\cal X}^0 =0$ for $M=m$, and
$({\cal X}^0 D')'=-{\cal C} D'$ for $M=0$. The remaining results
follow from (\ref{kill1}) and (\ref{kill2}).
To show this, we regard (\ref{kill1}) and (\ref{kill2}),
for any fixed function ${\cal X}^0(\varphi)$,
as a set of inhomogeneous equations for the ${\cal X}^m$, ${\cal A}_M$,
${\cal B}_M$ and ${\cal Y}_M$. The general solution is then the sum
of a particular solution and the general solution of the homogeneous
equations. A particular solution is given by
\begin{equation}
{\cal X}^{(p)m}=- (f'/2f) {\cal X}^0 x^m
\label{part}\end{equation} 
and corresponding expressions
for ${\cal A}^{(p)}_M$, ${\cal B}^{(p)}_M$ and ${\cal Y}^{(p)}_M$ 
obtained from
(\ref{sol}) for ${\cal X}^m\rightarrow{\cal X}^{(p)m}$.
The homogeneous equations (\ref{kill1}) and (\ref{kill2}), obtained
by setting ${\cal X}^0=0$, yield, for $(M,N)=(m,n)$, $(M,N)=(m,0)$ and
$(M,N)=(0,0)$ respectively,
\begin{eqnarray}
&& 
\eta_{nk}\partial_m {\cal X}^{(h)k}+\eta_{mk}\partial_n {\cal X}^{(h)k}=0,
\label{spec1}\\
&& \eta_{mn}({\cal X}^{(h)n})'f=-{\cal A}^{(h)}_m D',\quad 
0= {\cal A}^{(h)}_0 D', 
\label{spec2}\\
&&
\partial_n\tilde {\cal Y}^{(h)}_m-\partial_m\tilde {\cal Y}^{(h)}_n=0,
\label{spec3}\\
& &
(\tilde {\cal Y}^{(h)}_m)'-\partial_m{\cal Y}^{(h)}_0
-{\cal B}^{(h)}_m D'+B'_{mn}{\cal X}^{(h)n}=0
\label{spec4}\end{eqnarray}
where we used that $B_{mn}$ depends only on $\varphi$ and
defined
\[
\tilde {\cal Y}^{(h)}_m={\cal Y}^{(h)}_m-B_{mn}{\cal X}^{(h)n}\ .
\]
(\ref{spec1}) are just the Killing vector equations
for a flat space with coordinates $x^m$
and thus have the general solution
\begin{equation}
{\cal X}^{(h)m}=a^m(\varphi)+a^{mn}(\varphi)x_n,\ a^{mn}=-a^{nm}.
\label{hom}\end{equation}
(\ref{spec2}) can be solved for the ${\cal A}^{(h)}_M$ and thus
determines directly these functions.
(\ref{spec3}) implies $\tilde {\cal Y}^{(h)}_m=\partial_m{\cal Y}$ 
for some ${\cal Y}(X)$. Using this
in (\ref{spec4}), we get 
${\cal B}^{(h)}_m=\partial_m{\cal B}+B'_{mn}{\cal X}^{(h)n}/D'$ with
${\cal B}=({\cal Y}'-{\cal Y}^{(h)}_0)/D'$, and thus also
${\cal Y}^{(h)}_0 ={\cal Y}'-{\cal B} D'$. Furthermore, we
have the trivial identity
${\cal B}^{(h)}_0={\cal B}'+{\cal E} D'$ with 
${\cal E}=({\cal B}^{(h)}_0-{\cal B}')/ D'$.
Now, contributions $\partial_M{\cal Y}-{\cal B} \partial_M D$ and 
$\partial_M{\cal B}+{\cal E}\partial_M D$
to ${\cal Y}_M$ and ${\cal B}_M$ respectively can be removed by
redefinitions (\ref{triv}). Without loss of generality, 
we can thus choose
\begin{eqnarray}
& &
{\cal Y}^{(h)}_m=B_{mn}{\cal X}^{(h)n},\ {\cal Y}^{(h)}_0=0,
\nonumber\\
& &
{\cal B}^{(h)}_m=B'_{mn}{\cal X}^{(h)n}/D',\ {\cal B}^{(h)}_0=0.
\label{hom2}\end{eqnarray}
Altogether this yields (\ref{sol}).

Let us now discuss the symmetries (\ref{trafos})
arising from (\ref{sol}). As they involve arbitrary functions
${\cal X}^0(\varphi)$, $a^m(\varphi)$ and $a^{mn}(\varphi)$,
we conclude immediately
that any model characterized by (\ref{simple}) possesses
{\em infinitely many nontrivial rigid symmetries}.
To interpret them, we will use the equations of motion,
as rigid symmetries map in general solutions to 
the equations of motion to other solutions.
The equations of motion for $\gamma_{\mu\nu}$ are
solved for instance by $\gamma_{\mu\nu}={\cal G}_{\mu\nu}$
with ${\cal G}_{\mu\nu}$ as in (\ref{Sp}).
The equation of motion for $a_\mu$ yields 
\begin{equation} 
\epsilon^{\mu\nu}\partial_\nu D(\varphi)= 0\ \Rightarrow\ 
\varphi= \varphi_0=constant
\label{eoma}\end{equation}
i.e. $\varphi$ is on-shell just a constant fixed
by initial conditions. The value of this constant
distinguishes thus partly different solutions. Furthermore it
controls among others the coupling of the gauge field
to the string, as the equations of motion for $\varphi$ and 
$x^m$ yield respectively
\begin{eqnarray}
& &
\epsilon^{\mu\nu}(F_{\mu\nu}+{\cal B}_{\mu\nu})
= K(\varphi_0)\sqrt {\cal G}\ ,
\label{eomphi}\\
& &
\partial_\mu(\sqrt {\cal G}\, {\cal G}^{\mu\nu}\partial_\nu x_m+
\tilde B_{mn}(\varphi_0)\epsilon^{\mu\nu}\partial_\nu x^n)= 0
\label{eomx}\end{eqnarray}
where we have defined 
\begin{eqnarray*}
& &{\cal G}=-\det ({\cal G}_{\mu\nu}),
\\ 
& &{\cal B}_{\mu\nu}=B_{mn}(\varphi_0)\partial_\mu x^m\cdot\partial_\nu x^n,
\\
& &\tilde B_{mn}(\varphi_0)=B_{mn}(\varphi_0)D(\varphi_0)/f(\varphi_0),\\
& &K(\varphi_0)=-2f'(\varphi_0)/D'(\varphi_0).
\end{eqnarray*}
Note that (\ref{eomx}) are nothing but the equations of motion
for an ordinary bosonic string with constant $\tilde B_{mn}$ and
that (\ref{eomphi}) relates the abelian gauge field to this string.
Now, $\Delta x^m$ reads
\begin{equation}
\Delta x^m= a(\varphi)x^m+a^m(\varphi)+a^{mn}(\varphi)x_n
\label{Dex}\end{equation}
where $a=-{\cal X}^0f'/2f$. As $\varphi$ is constant for any solution
of the equations of motion, (\ref{Dex}) generates on-shell Poincar\'e 
transformations and dilatations\footnote{These 
dilatational symmetries should neither be confused
with the world-sheet Weyl invariance of (\ref{PS1}) and the world-volume
Weyl invariance of certain formulations of the (super) $p$-brane, nor with
the linearly realized global target space scale invariance of 
the formulation \cite{paul} treating the string tension of a $p$-brane
as a dynamical variable.}
of the target space coordinates.
The important property of the new symmetries is that they
transform in addition the gauge field $a_\mu$ nontrivially.
In particular, for a transformation
(\ref{Dex}) which generates on-shell
a dilatation of $x^m$, we get in the case $B_{mn}=0$:
\begin{eqnarray}
& & B_{mn}=0,\ \Delta x^m=a(\varphi)x^m
\quad \Rightarrow
\nonumber\\
& &
\Delta a_\mu=
\epsilon_{\mu\nu}
{\cal A}(\varphi)\sqrt {\cal G}\,{\cal G}^{\nu\rho}x^m\partial_\rho x_m
+{\cal C}(\varphi)\, a_\mu
\label{Dea}\end{eqnarray}
where ${\cal A}=a'f/D'$ and ${\cal C}=(2afD'/f')'/D'$.
Now, even for $a_\mu=0$, (\ref{Dea}) does in general
not reduce to a gauge transformation (not even on-shell!).
In particular, it
maps thus in general a solution to the equations of motion 
with $a_\mu=0$ to another solution with nonvanishing field
strength $F_{\mu\nu}$. Indeed, (\ref{eomphi}) shows that
solutions with vanishing $F_{\mu\nu}$ correspond in the
case $B_{mn}=0$ to special values of $\varphi_0$, namely
roots of the function $K$,
\[
B_{mn}=0:\  F_{\mu\nu}=0\ \leftrightarrow\ K(\varphi_0)=0.
\]
As transformations (\ref{Dea}) are accompanied by transformations
$\Delta \varphi={\cal X}^0(\varphi)$ (recall that $a=-{\cal X}^0f'/2f$),
we conclude: {\em in models with
$B_{mn}=0$, any transformation $\Delta \varphi={\cal X}^0(\varphi)$
which changes the value of $K(\varphi_0)$ from 0 to a nonvanishing
one, is accompanied by a transformation
$F_{\mu\nu}=0$ $\rightarrow$ $F_{\mu\nu}\neq 0$}!
Completely analogous considerations apply of course to models
with $B_{mn}\neq 0$.

Let us now discuss the off-shell algebra of the symmetries
arising from (\ref{sol}). As the transformations (\ref{Dex})
can be regarded as $\varphi$-dependent Poincar\'e transformations
and dilatations of the target space coordinates, their algebra 
will in any
basis be a Ka\v{c}--Moody version of the Weyl algebra.
A basis is obtained by choosing a
suitable basis for the functions ${\cal X}^0(\varphi)$, 
$a^m(\varphi)$ and $a^{mn}(\varphi)$ occurring in (\ref{sol}), 
adapted to the properties (e.g.\ boundary conditions, topology)
of the specific model one wants to study.
To give an explicit example,
we consider the case
\[
f(\varphi)=\exp(\varphi)
\]
and functions ${\cal X}^0(\varphi)$, 
$a^m(\varphi)$ and $a^{mn}(\varphi)$ which can be expanded in
integer powers of $\exp(\varphi)$, i.e.
\begin{eqnarray*}
{\cal X}^0(\varphi)&=&c_a e^{-a\varphi},\\
a^m(\varphi)&=&c^m_a e^{-a\varphi},\ 
a^{mn}(\varphi)=c^{mn}_a e^{-a\varphi}
\end{eqnarray*}
where the $c$'s are constant infinitesimal
transformation parameters indexed by $a\in$\,{\bf Z}, and
summation over $a$ is understood.
We now decompose $\Delta$ according to
\[
\Delta=c_a L^a+c^m_a P^a_m+\frac 12\, c^{mn}_a M^a_{mn}
\]
where $L^a$, $P^a_m$ and $M^a_{mn}=-M^a_{nm}$ are the
generators of rigid symmetries, the algebra of 
which we want to compute. 
(\ref{trafos}) and (\ref{sol}) yield
\[
\begin{array}{ll}
L^a\varphi=e^{-a\varphi}, & 
L^a x^m=-\frac 12 e^{-a\varphi}x^m,\\
P^a_m\varphi=0,& P^a_m x^n=e^{-a\varphi}\delta^n_m,\\
M^a_{mn}\varphi=0,& 
M^a_{mn}x^k=e^{-a\varphi} (\delta^k_{m} x_{n}-\delta^k_{n} x_{m}).
\end{array}
\]
Analogously one determines readily the transformations of $a_\mu$.
For instance, one gets
\begin{eqnarray*}
L^a a_\mu&=&
{\cal Z}^a(\varphi)\sqrt \gamma\, \epsilon_{\mu\nu}\partial^\nu (x_m x^m)
\\
& &
+{\cal Z}^a_{mn}(\varphi)x^n\partial_\mu x^m
+{\cal C}^a(\varphi)\, a_\mu
\end{eqnarray*}
where
\begin{eqnarray*}
{\cal Z}^a &=& a\, e^{(1-a)\varphi}/(4D'), 
\\
{\cal Z}^a_{mn} &=& [B''_{mn}-(1+a)B'_{mn}]e^{-a\varphi}/(2D'),
\\
{\cal C}^a &=& e^{-a\varphi}(a-D''/D').
\end{eqnarray*}

It is now very easy to compute the symmetry
algebra on the $X^M$. 
On the gauge field it is more involved, but
by means of the general arguments
given in the previous section one concludes that
the algebra coincides necessarily on all fields
up to gauge transformations and on-shell
trivial symmetries of the type occurring in (\ref{triv1}).
If the latter are present, the algebra is open. This turns out
to be the case in general. However, at least for
$B_{mn}=0$ the algebra closes off-shell even on $a_\mu$
and reads
\begin{eqnarray*}
[L^a,L^b] &=&(a-b)\,L^{a+b} ,
\\
{}[L^a,P^b_m] &=&(\frac 12-b)\, P^{a+b}_m ,
\\
{}[L^a,M^b_{mn}] &=&-b\, M^{a+b}_{mn} ,
\\
{}[P^a_m,P^b_n] &=&0,
\\
{}[M^a_{mn},P^b_k] &=&
\eta_{km}P^{a+b}_n  -(m\leftrightarrow n),
\\
{}[M^a_{mn},M^b_{pq}] &=&
2\eta_{p[m}M^{a+b}_{n]q}  -(p\leftrightarrow q).
\end{eqnarray*}
This is what we call a
Ka\v{c}--Moody version of the Weyl algebra.

Let us briefly point out an immediate generalization of the above
results to models characterized by
\begin{eqnarray}
G_{0M}=0,\ G_{mn}=f(\varphi)\, g_{mn}(x),\quad\quad
\nonumber\\
B_{0m}=0,\ B_{mn}=h(\varphi)\, b_{mn}(x),\ D=D(\varphi),
\label{lesssimple}
\end{eqnarray}
i.e.\ we allow now of curved target space metrics $g_{mn}(x)$
and $x$-dependent $B_{mn}$.
Suppose that $\{\zeta^m_A(x),{\cal Y}_{mA}(x)\}$ is
a basis of inequivalent solutions to
\begin{equation}
{\cal L}_{\zeta_A} g_{mn}=0,\ 
{\cal L}_{\zeta_A} b_{mn}= 2\partial_{[n}{\cal Y}_{m]A}
\label{killing}
\end{equation}
where two solutions are called equivalent if they differ only 
by ${\cal Y}_m\rightarrow {\cal Y}_m+\partial_m{\cal Y}$.
Then solutions to (\ref{kill1}--\ref{kill3}) are given by
\begin{eqnarray}
{\cal X}^m&=&c^A(\varphi)\, \zeta^m_A(x),\ {\cal X}^0=0,
\nonumber\\
{\cal Y}_m&=&c^A(\varphi)\, {\cal Y}_{mA}(x),\ {\cal Y}_0=0,
\nonumber\\
{\cal A}_m&=&-(f/D')g_{mn}({\cal X}^n)',\ {\cal A}_0=0,
\nonumber\\
{\cal B}_m&=&(h/D')({\cal Y}_m-b_{mn}{\cal X}^n)',\ {\cal B}_0={\cal C}=0
\label{hom3}\end{eqnarray}
where $c^A(\varphi)$ are arbitrary functions.
We conclude that any Killing vector field $\zeta_A(x)$ of the
target space
satisfying (\ref{killing}) gives rise to infinitely
many rigid symmetries of the model characterized
by (\ref{lesssimple}). The algebra of these
symmetries is a Ka\v{c}--Moody version of the Lie algebra
of the ${\cal L}_{\zeta_A}$ and 
generalizes the  Poincar\'e Ka\v{c}--Moody algebra
found above. It appears to depend on $g_{mn}$ and $b_{mn}$
whether there is a generalization
of the dilatational symmetries too.

\section*{Supersymmetric extensions}

One might wonder whether there are supersymmetric
extensions of the symmetry structure presented above.
We have not investigated this question
in detail by means of a cohomological analysis.
However we have found such extensions
in simple cases.
For instance, consider the Lagrangian
\begin{equation} 
L= 
\frac 12\, f(\varphi)\sqrt \gamma\, 
\gamma^{\mu\nu}\Pi_\mu^m\Pi_\nu^n\, \eta_{mn}
+D(\varphi)\epsilon^{\mu\nu}\partial_\mu a_\nu
\label{susy1}\end{equation}
where, as in (\ref{simple}), $f$ and $D\neq constant$ are any
functions of $\varphi$, and,
using the conventions and notation of \cite{actions2},
\begin{equation} 
\Pi_\mu^m=\partial_\mu x^m-\bar \theta\, \Gamma^m\partial_\mu\theta .
\label{Pi}
\end{equation}
The action with Lagrangian (\ref{susy1})
is invariant under the following rigid supersymmetry
transformations
\begin{eqnarray}
Q\theta^\alpha &=& 
c^\alpha B(\varphi),
\nonumber\\
Q x^m &=& \bar c\,\Gamma^m\theta B(\varphi),
\nonumber\\
Q a_\mu &=& 2 \bar c\,\Gamma^m \theta\,
\sqrt \gamma\, \epsilon_{\mu\nu} \Pi_m^\nu
B'(\varphi)f(\varphi)/D'(\varphi) ,
\nonumber\\
Q\varphi&=&Q\gamma_{\mu\nu}=0
\label{susy2}
\end{eqnarray}
where $c^\alpha$ is a constant anticommuting target
space spinor, $B(\varphi)$ is an arbitrary function of
$\varphi$, and 
\[\Pi_m^\mu=\eta_{mn}\gamma^{\mu\nu}\Pi^n_\nu\ .\]
The commutator of two transformations (\ref{susy2}) reads
\begin{equation}
[Q_1,Q_2]=2\, \bar c_2 \Gamma^m c_1\, P_m
\label{susy3}
\end{equation}
where $P_m$ generates $\varphi$-dependent ``translations''
of the type found above in the nonsupersymmetric case,
\begin{eqnarray}
P_m x^n &=&\delta_m^n\, B_{12}(\varphi)\ ,
\nonumber\\
P_m a_\mu &=& 
\sqrt \gamma\, \epsilon_{\mu\nu}\Pi_m^\nu
B'_{12}(\varphi)f(\varphi)/D'(\varphi),
\nonumber\\
P_m\varphi&=&P_m\gamma_{\mu\nu}=P_m\theta^\alpha=0
\label{susy4}
\end{eqnarray}
with
\[ B_{12}(\varphi)=B_1(\varphi)B_2(\varphi). \] 
Together with the analogues of the
symmetries arising from (\ref{trafos}), the 
supersymmetries (\ref{susy2}) form a Ka\v{c}--Moody
super-Weyl algebra which can be easily constructed
explicitly along the lines of the previous section.

\section*{Conclusion}

Any local action in two
dimensions ($p=1$) with field content
and gauge symmetries given by (\ref{fields}) and
(\ref{brs}) respectively, has a Lagrangian of the form 
(\ref{L}). Specific choices of $G_{MN}$, $B_{MN}$ and $D_I$
provide actions which turn
upon elimination of the auxiliary fields
into $D$-string actions of the
Born--Infeld type such as (\ref{Sp}) (for $p=1$) or,
more generally, (\ref{BI2}) or (\ref{BI3}).

The rigid symmetries of an action with Lagrangian
(\ref{L}) are determined by the solutions of the
generalized Killing vector equations (\ref{kill1}--\ref{kill3}).
We have shown that these equations can have
infinitely many inequivalent solutions which we have
spelled out explicitly for specific models in a
flat target space. In the latter models,
we have found a Ka\v{c}-Moody realization of the Weyl group, the new 
symmetries being non-linearly realized. 
Symmetries of the actions (\ref{Sp}) (for $p=1$),
(\ref{BI2}) and (\ref{BI3}) are obtained from those of their
counterparts (\ref{L}) simply by eliminating the auxiliary fields.
For instance, from (\ref{Dea}) one obtains in this way among others
a symmetry of the $p=1$-action (\ref{Sp}) generated by
\begin{eqnarray}
\Delta x^m&=&F\, x^m,
\nonumber\\
\Delta a_\mu&=&
\epsilon_{\mu\nu}
(F^2-1)\sqrt {\cal G}\, {\cal G}^{\nu\rho}x^m\partial_\rho x_m
+2F\, a_\mu
\label{dils}\end{eqnarray}
where, assuming that ${\cal G}_{\mu\nu}$ has Lorentzian signature,
\[ F={\cal G}^{-1/2}\, \epsilon^{\mu\nu}\partial_\mu a_\nu\, ,\quad
{\cal G}=-\det ({\cal G}_{\mu\nu}).\]
$F$ is constant on-shell. The value of this
constant characterizes partly a solution
to the equations of motion and contributes to its string
tension. (\ref{dils}) generates on-shell a
dilatation of the
target space coordinates, but it also transforms 
the abelian gauge field nontrivially. In particular, it
transforms a solution to the equations of motion
with $F=0$ to another one with $F\neq 0$, as on-shell one
has $\Delta F=2(F^2-1)$.
Symmetries such as (\ref{dils}) are thus
useful, among others, to connect configurations of the 
fundamental string with those of the D-string.

We have also
shown that the Ka\v{c}--Moody symmetry structure 
extends analogously to curved target spaces
if the latter possess Killing vector fields satisfying
(\ref{killing}). Furthermore we have given some examples of
supersymmetric extensions where an infinite number of rigid
supersymmetries appears in addition to the Ka\v{c}--Moody--Weyl
symmetries.

We have obtained our results by an analysis of the BRST cohomology
at ghost numbers 0 (actions) and $-1$ (symmetries). 
Especially the results on the rigid symmetries are difficult to
guess or to derive by other means due to highly nonlinear
nature of symmetries such as (\ref{dils}). 

One may speculate whether the
infinite number of symmetries reflects part of 
the space-time symmetry structure of an underlying (``M'') theory. 
This possibility is suggested because $D$-branes may probe 
shorter space-time distances than strings \cite{shenker}.
It would be interesting to further understand the physical meaning 
of the Weyl--Ka\v{c}--Moody algebra and whether or not
the $\kappa$-invariant formulation of the $D$-string
has infinitely many rigid symmetries too.
Another interesting point to be investigated 
will be to check whether a Weyl--Ka\v{c}--Moody algebra 
appears also for other $D$-$p$-branes.

\section*{Acknowledgements}

J. Gomis acknowledges Kiyoshi Kamimura for discussions on $D$-brane
actions. 
J. Sim\'on thanks Fundaci\'o Agust\'{\i} Pedro i Pons 
for financial support.
F. Brandt was supported by the Spanish ministry of education and 
science (MEC).
This work has been partially supported by AEN95-0590 (CICYT),
GRQ93-1047 (CIRIT) and by the 
Commission of European Communities CHRX93-0362 (04).

\end{document}